\documentclass[11pt]{article}
\usepackage{moriond,epsfig}

\def\be{\begin{equation}}
\def\ee{\end{equation}}
\def\bea{\begin{eqnarray}}
\def\eea{\end{eqnarray}}

\newcommand{\tetaot}{\mbox{$\theta_{13}$}}

\newcommand{\tetatt}{\mbox{$\theta_{23}$}}

\begin{document}
\begin{flushright}
{FTUAM-03-08}\\
{IFT-UAM/CSIC-03-14} \\
{Presented at XXXVIII Rencontre De Moriond:}\\
{Electroweak Interactions and Unified Theories}\\
{Les Arcs, France, March 15--22, 2003}
\end{flushright}

\vspace*{2cm}
\title{PUZZLING OUT NEUTRINO MIXING THROUGH GOLDEN AND SILVER MEASUREMENTS}

\author{O. Mena}

\address{Dept. de F\'{\i}sica Te\'orica, Univ. Aut\'onoma de
Madrid, Spain}
    
\maketitle\abstracts{We update a recent work devoted to resolve the degeneracies that appear in the simultaneous extraction of $\theta_{13}$ and $\delta$ at future Neutrino Factories (NF, that exploit the \emph{golden} channels, i.e. $\nu_e\rightarrow\nu_\mu$ ($\bar{\nu}_e\rightarrow\bar{\nu}_\mu$)) and Superbeam experiments (SB, that measure the $\nu_\mu\rightarrow\nu_e$ ($\bar{\nu}_\mu\rightarrow\bar{\nu}_e$) transitions). We consider the neutrino fluxes obtained with a new optics design for the CERN-SPL SB and assume the solar parameters within the LMA-I and the LMA-II regions indicated by recent KamLAND data. The dangerous fake solution associated with the $\theta_{23}$-ambiguity remains after the combination of data from these two facilities: in this perspective, we analyze the impact of the NF-\emph{silver} channels, i.e. $\nu_e\rightarrow\nu_\tau$ ($\bar{\nu}_e\rightarrow\bar{\nu}_\tau$). The combination of data from these three experiments-NF(\emph{golden} and \emph{silver} channels) plus SPL SB- can discover leptonic CP violation for values of $\theta_{13}\ge 1^{\circ}$.} 

\section{Introduction}

The neutrino puzzle starts to be disentangled: the reactor experiment KamLAND~\cite{kam} has confirmed the LMA-MSW~\cite{MSW} solution . Two possible regions (called LMA-I and LMA-II) have been identified in the two family oscillation scenario~\cite{lisik,concha,otros} after combining KamLAND, CHOOZ and solar data. As it was pointed out in Ref.~\cite{golden}, this is the only solution of the solar neutrino problem for which it could be possible to be sensitive to the leptonic CP-violating phase $\delta$, provided that $\delta$ and $\tetaot$ are not too small.

Notice that the measurement of $\delta$ must be performed at the same time as that of $\tetaot$, being the latter a subleading parameter in the processes studied at the existing and planned experiments. The correlations between $\delta$ 
and $\tetaot$ give rise to undesired \emph{degeneracies} in the parameter space~\cite{burguet,lisi,firstsign,bargerdeg}, similar to the solar neutrino data previous than KamLAND and SNO experiments.

A lot of work has been carried on in the past two years in order to eliminate some of these fake solutions, advocating the combination of 
different baselines~\cite{burguet}; an improved experimental technique allowing the measurement of the neutrino 
energy with good precision~\cite{lindner}; the supplementary detection of $\nu_e \rightarrow \nu_\tau$ transitions~\cite{andrea}; or the combination of a cluster of detectors at
a superbeam facility located at different angles with respect to the beam axis, so as to have 
different $\langle E\rangle$~\cite{bargernew}. The authors of Ref.~\cite{nuno} performed an analytic study of the parameter degeneracies and more recently they have considered different runnings ($\nu_\mu$ or $\bar \nu_\mu$) at JHF~\cite{jhf} and NuMI Off Axis~\cite{oa} experiments in order to extract the sign of $\Delta m^{2}_{13}$ and/or $\tetaot$, see Ref.~\cite{parke}. A detailed analysis of different suitable detectors for the NuMI Off Axis beam option has been recently presented~\cite{fermi}.   

\section{KamLAND data plus SPL new optics design}

 It was shown in Ref.~\cite{nuevo} how almost all the degeneracies were eliminated by combining the expected results from NF and
 CERN-SPL SB due to the different $E_{\nu}$, $L$, dependence of the two experiments. However, in the previous numerical analysis, a rather optimistic value for $\Delta m^{2}_{12}$ was considered, $\Delta m^{2}_{12}=1 \times 10^{-4}$ eV$^{2}$ and $\sin^{2} 2 \theta_{12}=1$.

The best fit values in Ref.~\cite{lisik} for LMA-I -the most favored solar parameter region after KamLAND data- are $\Delta m^{2}_{12}=7.3 \times 10^{-5}$ eV$^{2}$ and $\sin^{2} 2 \theta_{12}=0.86$. In principle, this decrease of $40\%$ in $\Delta m^{2}_{12}\cdot\sin^{2} 2 \theta_{12}$ could have important consequences with respect to the conclusions in Ref.~\cite{nuevo}, as regarding the extraction of $\tetaot$ and $\delta$. 

\begin{figure}
\begin{center}
\mbox{
\psfig{figure=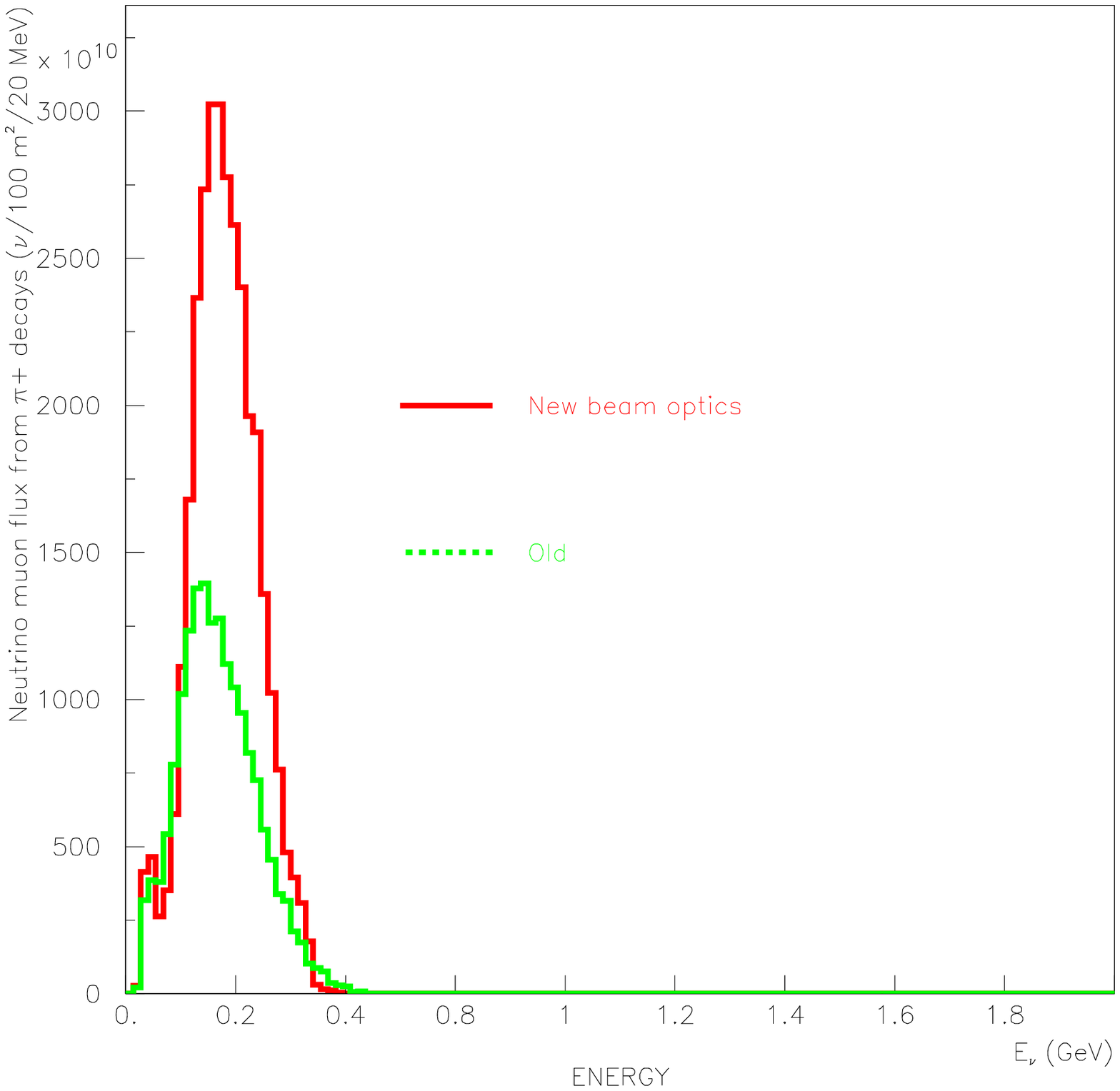,height=2.5in}
\psfig{figure=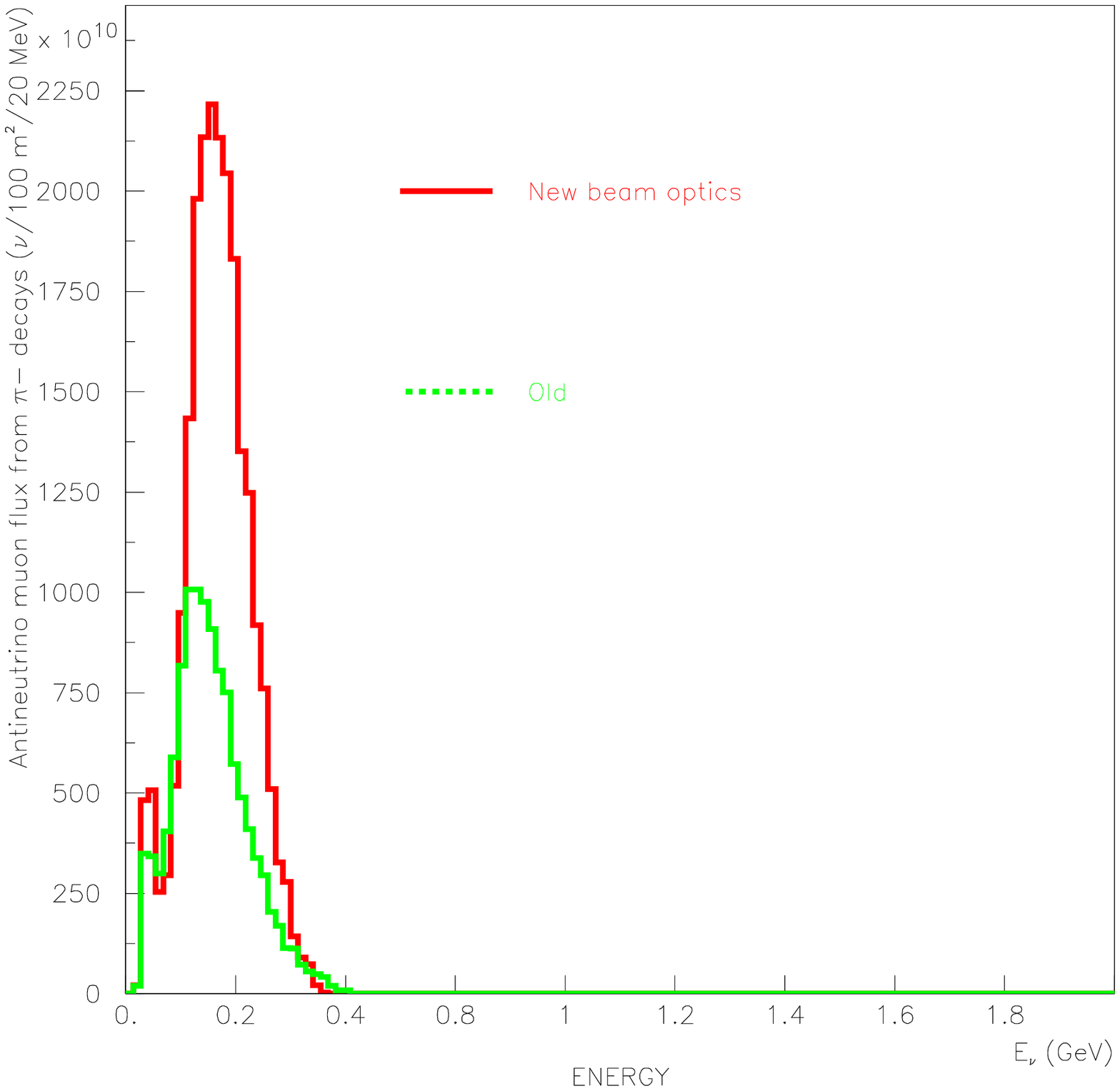,height=2.5in}}
\end{center}
\caption{The CERN-SPL neutrino (left) and antineutrino (right) spectra
for $\pi^+$ (left) and for $\pi^-$ (right) focused in the horn. We show the
beam with the new optics design vs. old beam.
\label{fluxes}}
\end{figure}

\begin{figure}[h]
\begin{center}
\mbox{
\psfig{file=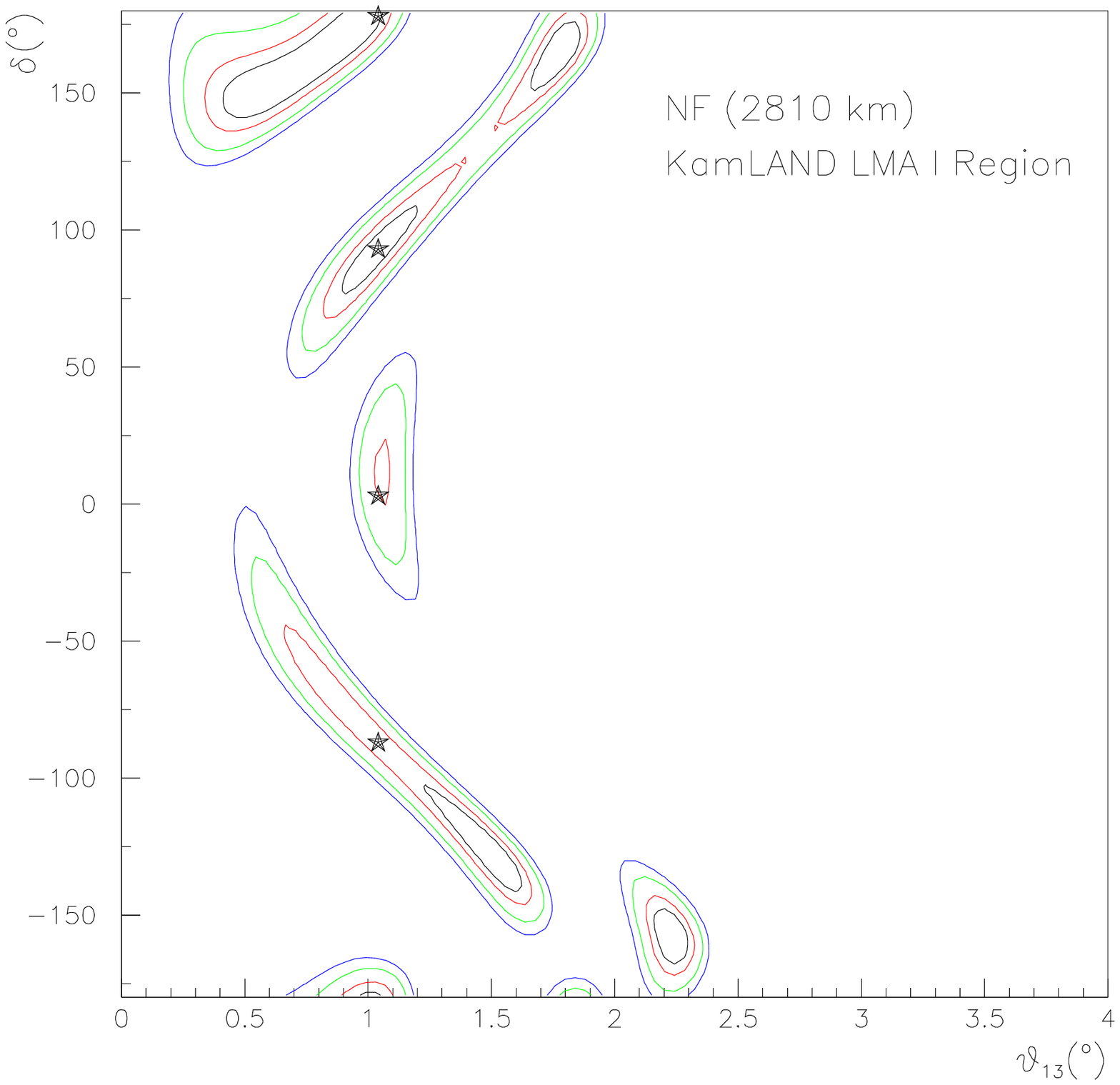,height=3in}
\psfig{file=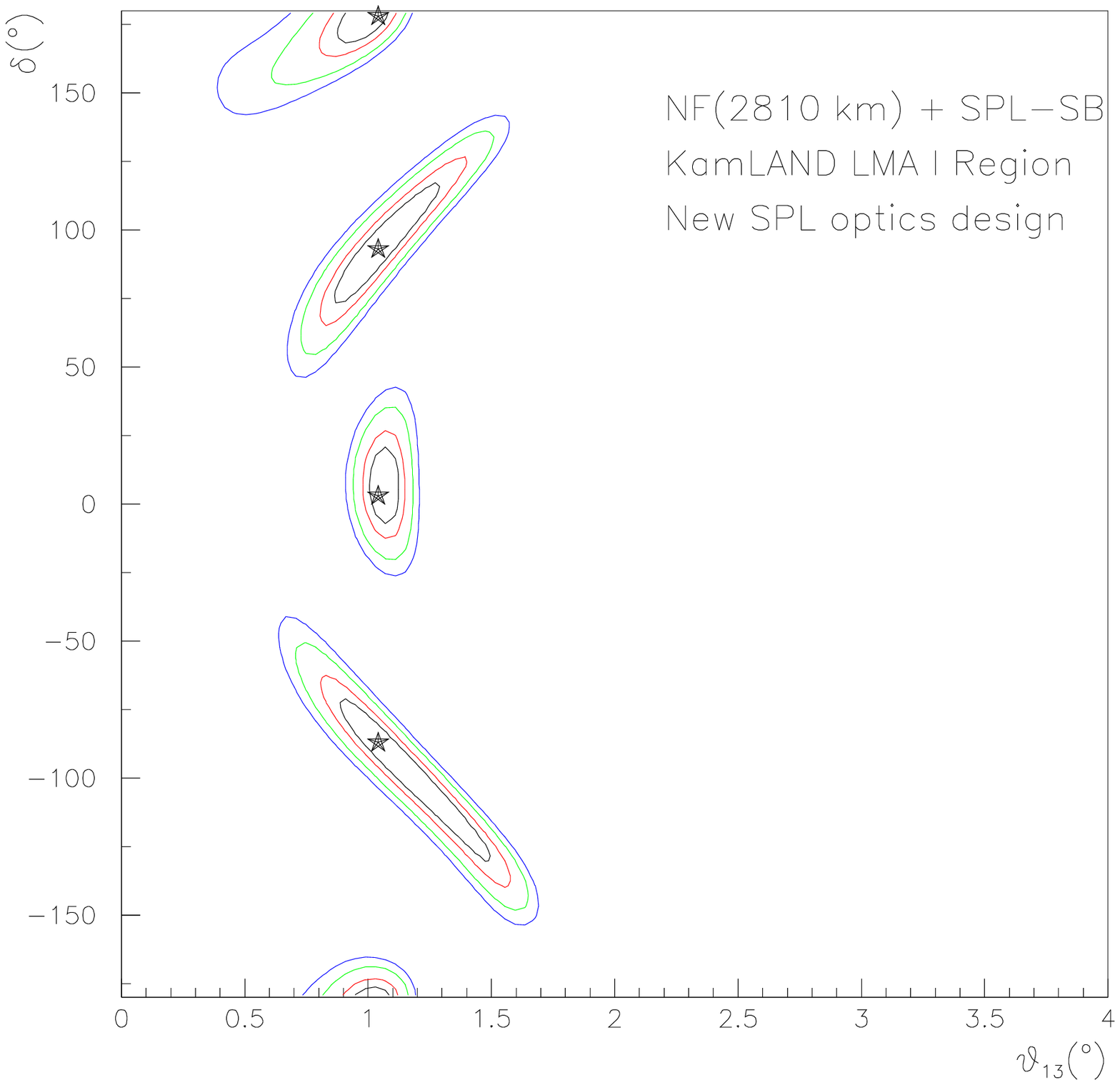,height=3in}
}
\end{center}
\caption{1, 2, 3 and 4$\sigma$ contours resulting from the fits  
before and after combining the results from a NF baseline at $L=2810$ km 
with those from the CERN-SPL facility.
The true values illustrated (indicated by a star) correspond to 
 $\delta=- 90^\circ, 0^\circ, 90^\circ, 180^\circ$ and $\tetaot=1^\circ$.
\label{sim}}
\end{figure}

Fortunately, a new optics design for the CERN-SPL SB has been recently made available~\cite{simone} and we shall consider it here. The neutrino beam from the new optics design is much more intense than the one in the first CERN-SPL proposal~\cite{sb2,sb}, as can be noted in Fig.~\ref{fluxes}.(In these references the detector systematic was also discussed).  
For the NF we consider the experimental setup presented
 in Refs.~\cite{golden,burguet,nuevo,lmd}. Regarding the neutrino oscillation parameters, whenever it is not specified we take $|\Delta m^2_{13}| = 3\times 10^{-3}$ eV$^2$ and $\sin 2 \theta_{23} = 1$, and the solar parameters within the LMA-I region. Fig.~\ref{sim} shows the results of measuring $(\tetaot,\delta)$ before and after the combination of the data from a NF baseline at $L=2810$ km with the data from the CERN-SPL facility for $\tetaot = 1^{\circ}$ and the central values 
of $\delta=-180, -90, 90, 180^\circ$.  Notice that fake \emph{intrinsic} solutions have completely disappeared. For larger values of $\tetaot$ the fit shows  much smaller uncertainties on the determination of the CP-violating phase $\delta$.

\begin{figure}
\begin{center}
\mbox{
\psfig{file=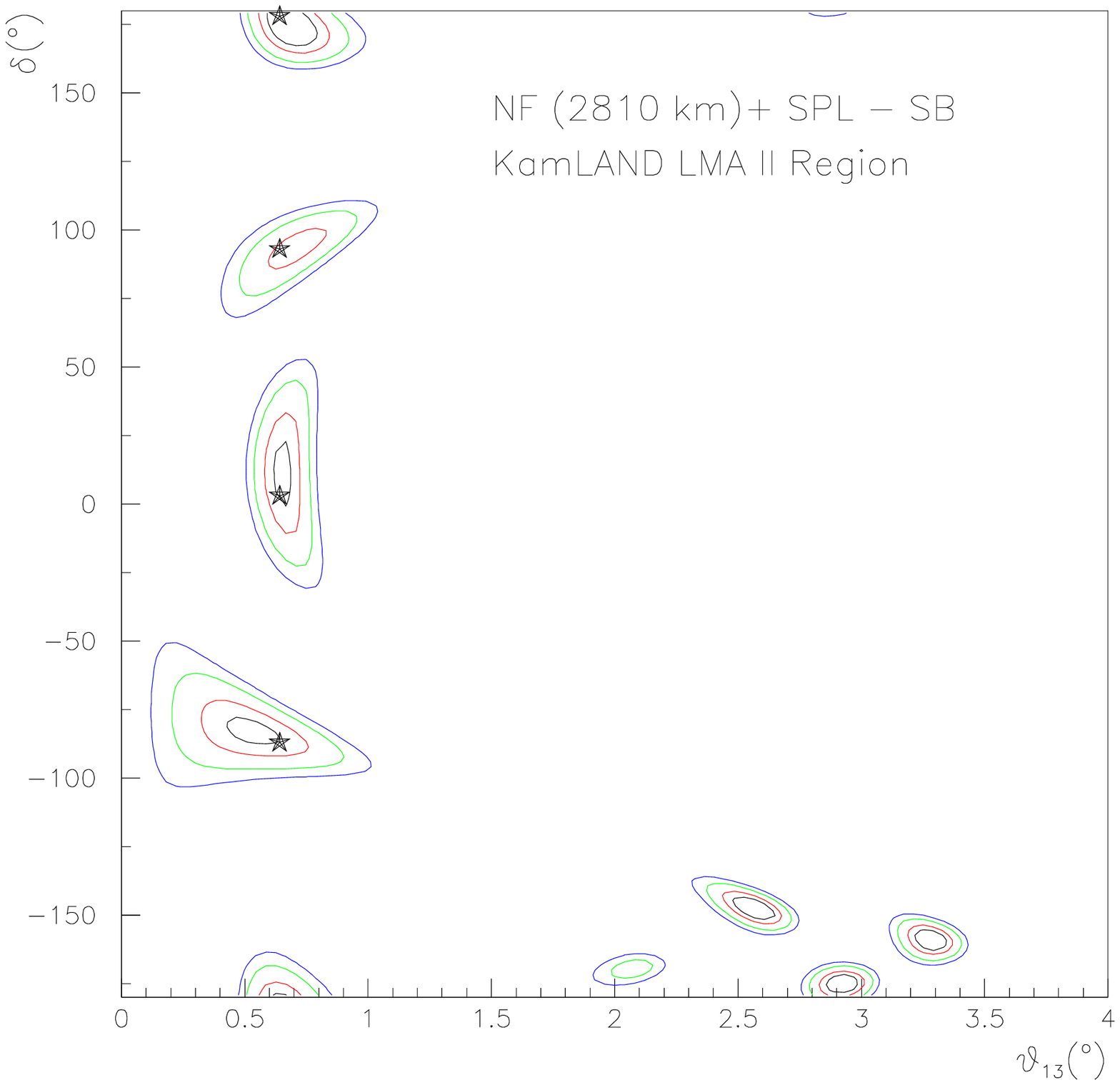,height=3in}
\psfig{file=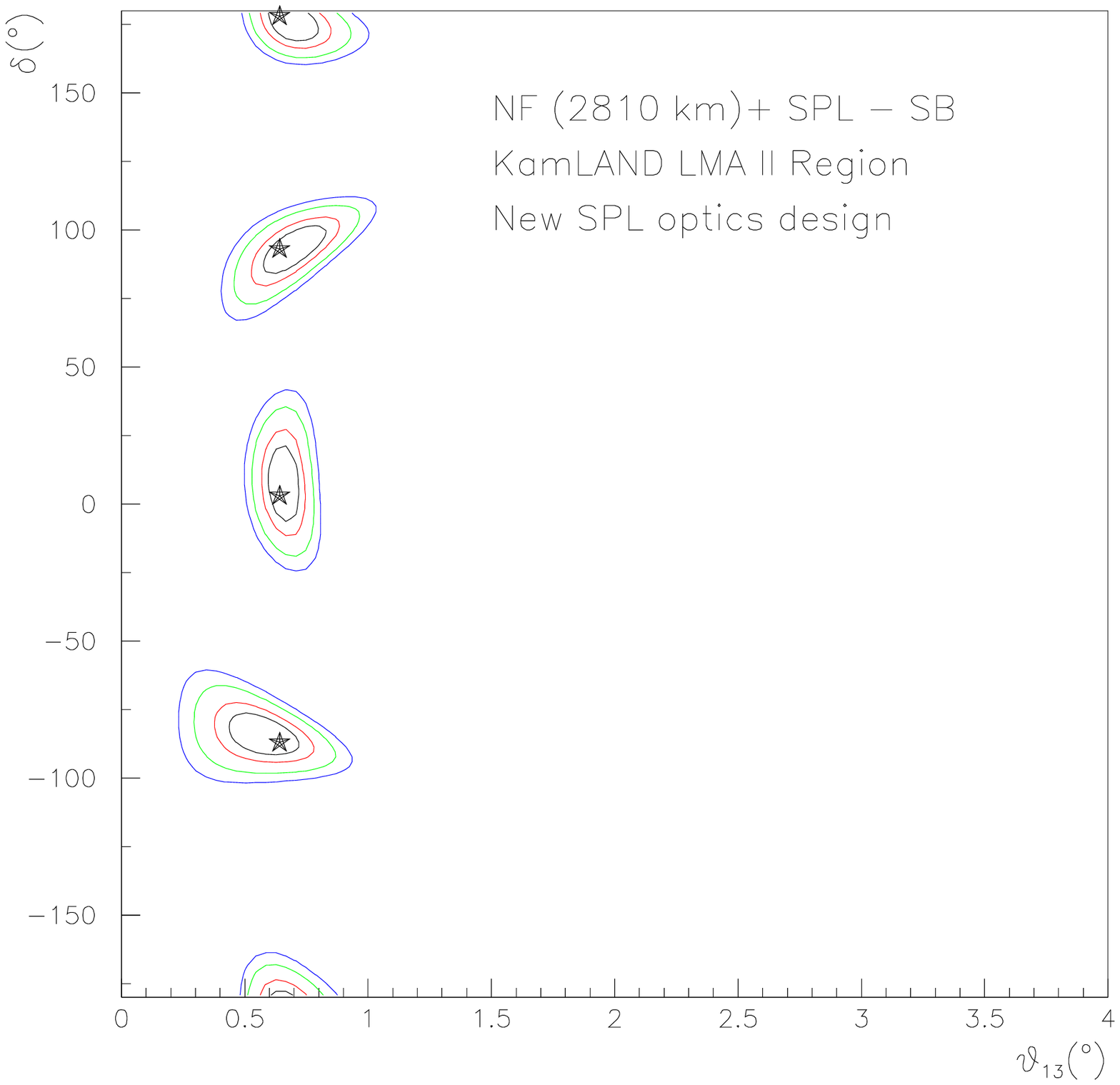,height=3in}
}
\end{center}
\caption{The same as Fig.~\ref{sim} but for the solar parameters within the LMA-II region and $\tetaot=0.6^\circ$.
\label{sign}}
\end{figure}

We have also considered the solar parameters within the LMA-II region~\cite{lisik}: $\Delta m^{2}_{12}=1.54\times 10^{-4}$ eV$^{2}$, $\sin^{2} 2\theta_{12}=0.84$. These new values imply an improvement of $40\%$ respect to our previous central values~\cite{nuevo}.  In fact, within this region, we have found no \textit{intrinsic} degeneracies for values of $\tetaot > 0.6^\circ$, when considering one NF baseline at $L=2810$ km and CERN-SPL SB (see Fig.~\ref{sign}).

\section{$\tetatt$-octant ambiguity}
\begin{figure}
\begin{center}
\psfig{file=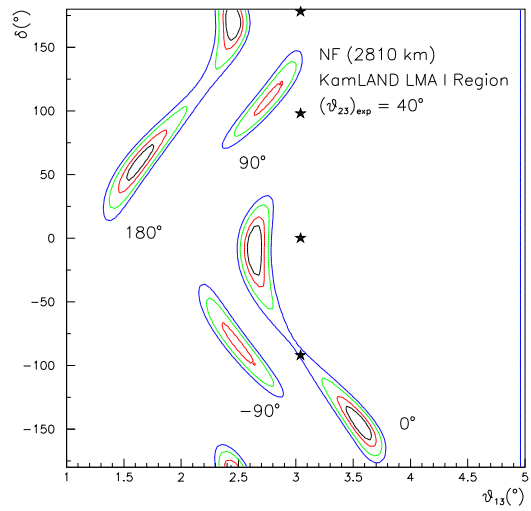,height=3in}
\end{center}
\caption{Fake solutions due to $\tetatt$ degeneracies for NF ($L=2810$ km) results, with the wrong choice of octant for $\tetatt$ ($\tetatt=40^{\circ}$), for  $\tetaot=3^{\circ}$ and $\delta=- 90^\circ, 0^\circ, 90^\circ, 180^\circ$. The combination of these results with those from a NF ($L=732$ km) exploiting the silver channels resolves the degeneracies.
\label{th23}}
\end{figure}

In previous work, Ref.~\cite{nuevo}, two types of fake solutions were found: those closer to nature's values (Solution I), $E/L$-independent, and consequently, difficult to overcome, and solutions of type II,  $E/L$-dependent. In spite of the new fluxes for the SPL-SB facility, the dangerous fake Solution I associated with the $\tetatt$-octant ambiguity remains after the combination of these data with those from a future NF (see Fig.~\ref{th23}): the extraction of $\tetaot$ and $\delta$ is not possible due to the location of the degeneracies.

However, as pointed out in Ref.~\cite{andrea}, an additional measurement of an independent appearance channel, $\nu_e\rightarrow\nu_\tau$  and $\bar{\nu}_e\leftrightarrow\bar{\nu}_\tau$ (silver channels) helps enormously to resolve the fake solution I associated with the $\tetatt$-ambiguity, since the locations of the fake solutions that arise from the data analysis of \emph{silver} and \emph{golden} channels differ substantially.

\begin{figure}
\begin{center}
\psfig{file=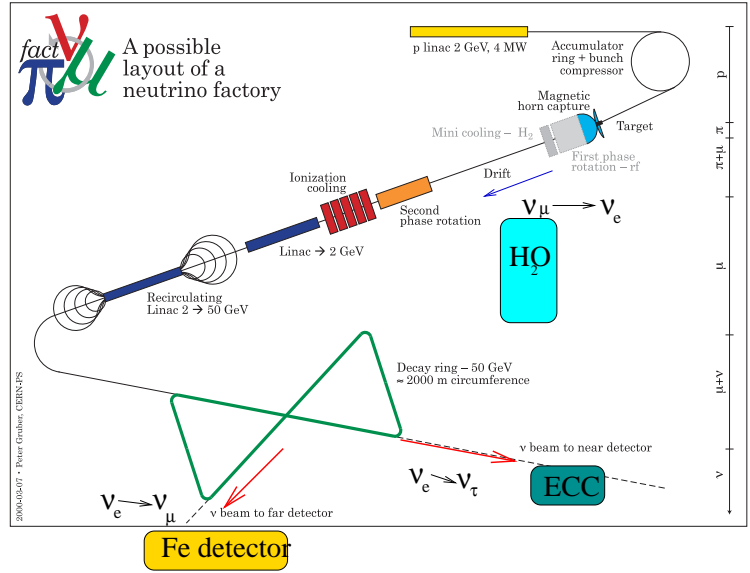,height=3in}
\end{center}
\caption{The complex of NF (golden and silver) and the SPL-SB experiments.
\label{schedule}}
\end{figure}

We have thus considered~\cite{prep} the impact of three simultaneous experiments (see Fig.~\ref{schedule}) with 2 years running in the $\pi^{+}$ polarity and 10 years running in the $\pi^{-}$ polarity. For the analysis of the silver channel we assume the setup of the Opera proposal~\cite{opera}, with one fixed baseline, $732$ km, i.e. the distance from CERN to CNGS, and a 4 Kton lead emulsion detector with spectrometers. A dedicated analysis can be found in Ref.~\cite{autiero}.

As the first stage of a detailed study with a realistic experimental setup~\cite{prep} we consider here an ideal situation, neglecting backgrounds and efficiencies for an emulsion cloud chamber (ECC) detector. We have found that after the combination of the results from NF golden ($L=2810$ km) and silver ($L=732$ km)  channels no degeneracy related to the $\tetatt$-octant ambiguity survives for $\tetaot>0.6^\circ$. This limit is expected to become more stringent, namely, $\tetaot>1^\circ$, when adding efficiencies and backgrounds. At very small values of $\tetaot$ ($\tetaot<0.6^\circ$), the $\tetatt$-octant ambiguity remains, but it does not interfere with the extraction of the two unknown parameters, $\tetaot$ and $\delta$, due to the location of the fake solution: $\tetaot^{'}\sim\tetaot$ and $\delta^{'}\sim \pi -\delta$.

\section{Summary}

After KamLAND data, the discovery of leptonic CP violation may be clearly at reach. Much research has been devoted to study the possibility for LBL experiments to measure precisely the neutrino oscillation parameters and hopefully the CP-odd phase $\delta$. 

We have shown the enormous potential of combining the data from the CERN-SPL SB and NF facilities to eliminate the degeneracies in the simultaneous measurement of $\delta$ and $\tetaot$ with neutrino fluxes from a new beam optics design for the CERN-SPL SB, which
 could be the first step of a NF based at CERN. We have considered both LMA-I and LMA-II regions for the solar parameters, the only that survive after KamLAND data. Previous results~\cite{nuevo} are unchanged in the case of the most favored solution, that is, the LMA-I, in spite of the notorious reduction of the solar parameters, due to the increased statistics.
 
The only degeneracy that survives after the combination is that associated with the $\tetatt$ ambiguity. We have shown that it is possible to eliminate it through the combination of NF golden and silver channels down to values of $\tetaot\sim 0.6^\circ$ in the present analysis. A realistic experimental scenario is under study~\cite{prep}.

It is very important to notice that there exist several alternative experimental setups that could help enormously in disentangling the neutrino puzzle: NuMI Off Axis beam, JHF~\cite{parke,fermi}, reactor experiments~\cite{huber} and the beta-beams facility~\cite{zucheli,mori}, whose potential is not discussed here. 

All the latter experimental options have to be thoroughly explored in order to ascertain the ultimate precision in the determination of the detailed pattern of neutrino mass differences and mixing angles, a prerequisite to understand their origin and their relationship to the analogous parameters in the quark sector. The NF (golden \textit{and} silver channels) plus its predecessor, the SB experiment, would provide the key to fulfill this goal.

\end{document}